\documentclass{article}

\usepackage{microtype}
\usepackage{graphicx}
\usepackage{subfigure}
\usepackage{booktabs} 

\usepackage{amssymb}

\usepackage{hyperref}

\usepackage[accepted]{icml2021}

\icmltitlerunning{Synthetic COVID-19 Chest X-ray Dataset for Computer-Aided Diagnosis}

\begin{document}

\twocolumn[
\icmltitle{Synthetic COVID-19 Chest X-ray Dataset for Computer-Aided Diagnosis}



\icmlsetsymbol{equal}{*}

\begin{icmlauthorlist}
\icmlauthor{Hasib Zunair}{ed}
\icmlauthor{A.~Ben Hamza}{ed}
\end{icmlauthorlist}

\icmlaffiliation{ed}{Concordia University, Montreal, QC, Canada}

\icmlcorrespondingauthor{Hasib Zunair}{hasibzunair@gmail.com}

\icmlkeywords{Machine Learning, ICML}

\vskip 0.3in
]



\printAffiliationsAndNotice{}  

\begin{abstract}
We introduce a new dataset called Synthetic COVID-19 Chest X-ray Dataset~\footnote{\href{https://github.com/hasibzunair/synthetic-covid-cxr-dataset}{\texttt{\textcolor{red}{GitHub:~Synthetic COVID-19 CXR Dataset}}}} for training machine learning models. The dataset consists of 21,295 synthetic COVID-19 chest X-ray images to be used for computer-aided diagnosis. These images, generated via an unsupervised domain adaptation approach, are of high quality. We find that the synthetic images not only improve performance of various deep learning architectures when used as additional training data under heavy imbalance conditions (skew $\geqslant$ 90), but also detect the target class with high confidence. We also find that comparable performance can also be achieved when trained only on synthetic images. Further, salient features of the synthetic COVID-19 images indicate that the distribution is significantly different from Non-COVID-19 classes, enabling a proper decision boundary. We hope the availability of such high fidelity chest X-ray images of COVID-19 will encourage advances in the development of diagnostic and/or management tools.
\end{abstract}

\section{Introduction} \label{Introduction}
Recent studies have shown that chest radiography images such as chest X-rays (CXR), performed on patients with COVID-19 when they arrive at the emergency room, can help doctors determine who is at higher risk of severe illness and intubation~\cite{ai2020correlation,huang2020clinical}. Automatic interpretation of chest radiography images such as CXR using computational approaches not only helps healthcare organizations save time and money, but also provides superior patient care and more importantly it can save lives~\cite{ng2020imaging}.

\begin{figure}[!htb]
\centering
\includegraphics[scale=.32]{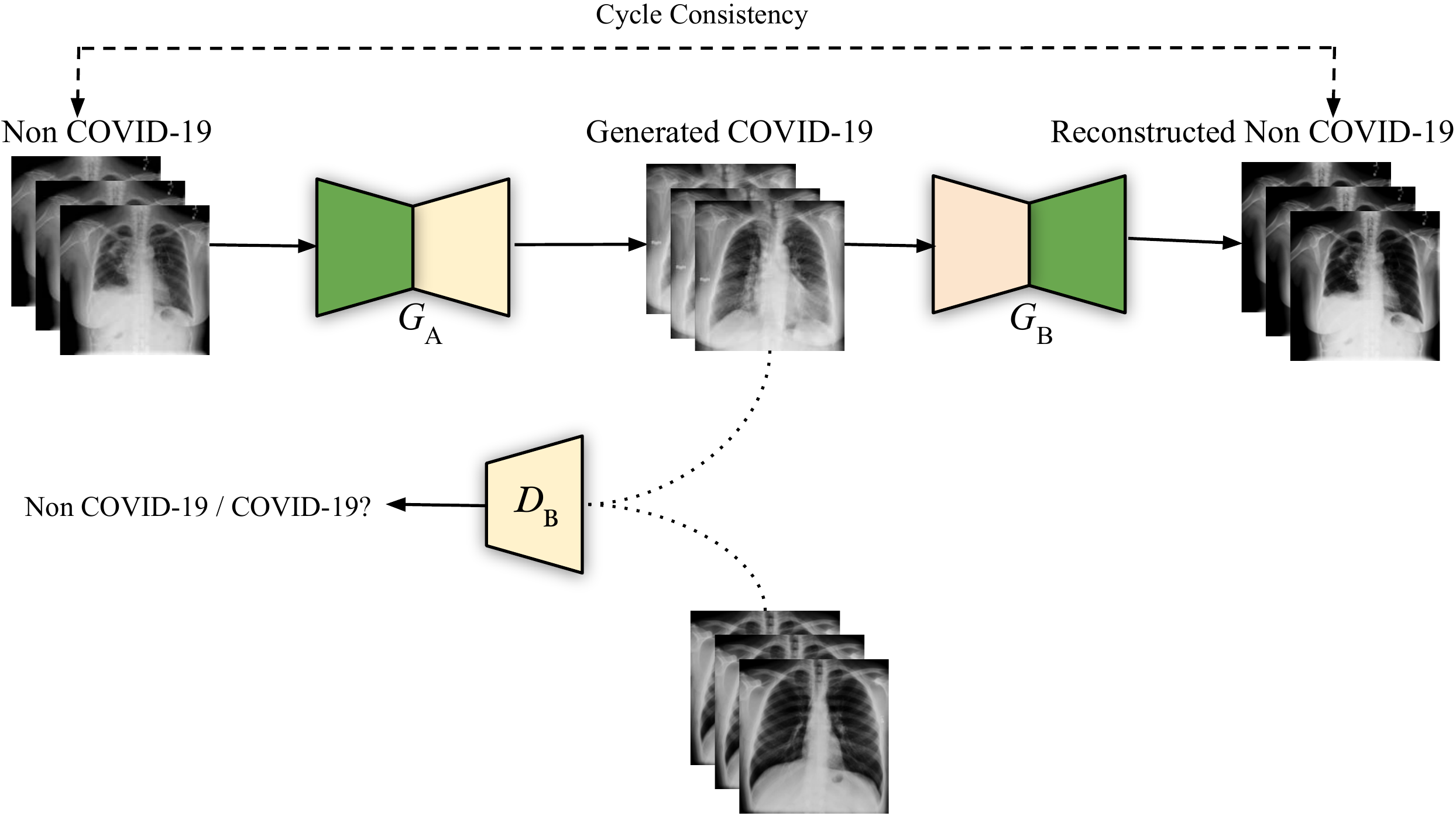}
\caption{Illustration of the data generation process based on unpaired image-to-image translation. CXR images are translated from Non-COVID-19 (i.e. Normal or Pneumonia) to COVID-19 and then back to Non-COVID-19 via cycle-consistency.}
\label{Fig:architecture}
\end{figure}

Several computational approaches for the detection of COVID-19 cases from chest radiography images have been recently proposed, including tailored convolutional neural network (CNN) architectures~\cite{karim2020deepcovidexplainer,wang2020covid} and transfer learning based methods~\cite{kassani2020automatic,narin2021automatic,XinLi:20,Farooq:20}. While promising, the predictive performance of these deep learning based approaches depends heavily on the availability of large amounts of curated and annotated data. Over the past few years, there have been several efforts to build large-scale
annotated datasets for CXRs and make them publicly available to the global research community~\cite{johnson2019mimic,wang2017chestx,bustos2020padchest}. At
the time of writing, there exists, however, only one annotated
COVID-19 X-ray Image Data Collection~\cite{cohen2020covid}, which is a curated collection of CXR images of patients who are positive or suspected of COVID-19 or other viral and bacterial pneumonia. This is largely attributed to the rare nature of the radiological finding, legal, privacy, technical, and data-ownership challenges. While the COVID-19 Image Data Collection contains positive examples of COVID-19, the negative examples were acquired from publicly available sources~\cite{wang2017chestx} and merged together for computational analysis. This fusion of multiple datasets results in predominantly negative examples with only a small percentage of positive ones (i.e COVID-19), giving rise to a class imbalance problem~\cite{johnson2019mimic,wang2017chestx,bustos2020padchest}.

\section{Methods}
\label{Methods}
Identifying COVID-19 in a CXR image can be regarded as an image classification problem. We consider two binary classification tasks, namely Normal vs. COVID-19 and Pneumonia vs. COVID-19, for which we want to accurately identify COVID-19. For training the classification and the generative models, we use COVID-19 samples from the COVID-19 Image Data Collection~\cite{cohen2020covid}, and Normal and Pneumonia samples from the RSNA Pneumonia Detection Challenge~\cite{wang2017chestx}.

In order to address the class imbalance problem, we recently proposed an unsupervised domain adaptation algorithm to synthesize under-represented class samples (i.e COVID-19 CXR images) from the over-represented ones (i.e Normal or Pneumonia CXR images) using unpaired image-to-image translation~\cite{zunair2021synthesis,zunair2020melanoma}. The key idea is to train a cycle-consistent generative model~\cite{zhu2017unpaired} on unpaired samples of two domains (Non-COVID-19 and COVID-19 in our case) with the goal of learning mapping functions between them. We train two translation models, which learn the mapping from Normal to COVID-19 and Pneumonia to COVID-19. After training, given a Non-COVID-19 CXR image (i.e. Normal or Pneumonia) as input, the generative model translates such that the image has representative features of COVID-19. An illustration of the data generation process in shown in Figure~\ref{Fig:architecture}. 

We generate 16,537 and 4,758 COVID-19 CXR images for Normal vs. COVID-19 (denoted $\mathcal{G}_{NC}$) and Pneumonia vs. COVID-19 (denoted $\mathcal{G}_{PC}$) tasks, respectively.

\section{Results}
\label{Results}
\begin{figure}[!htb]
\setlength{\tabcolsep}{-.1em}
\centering
\begin{tabular}{c}
\includegraphics[scale=.45]{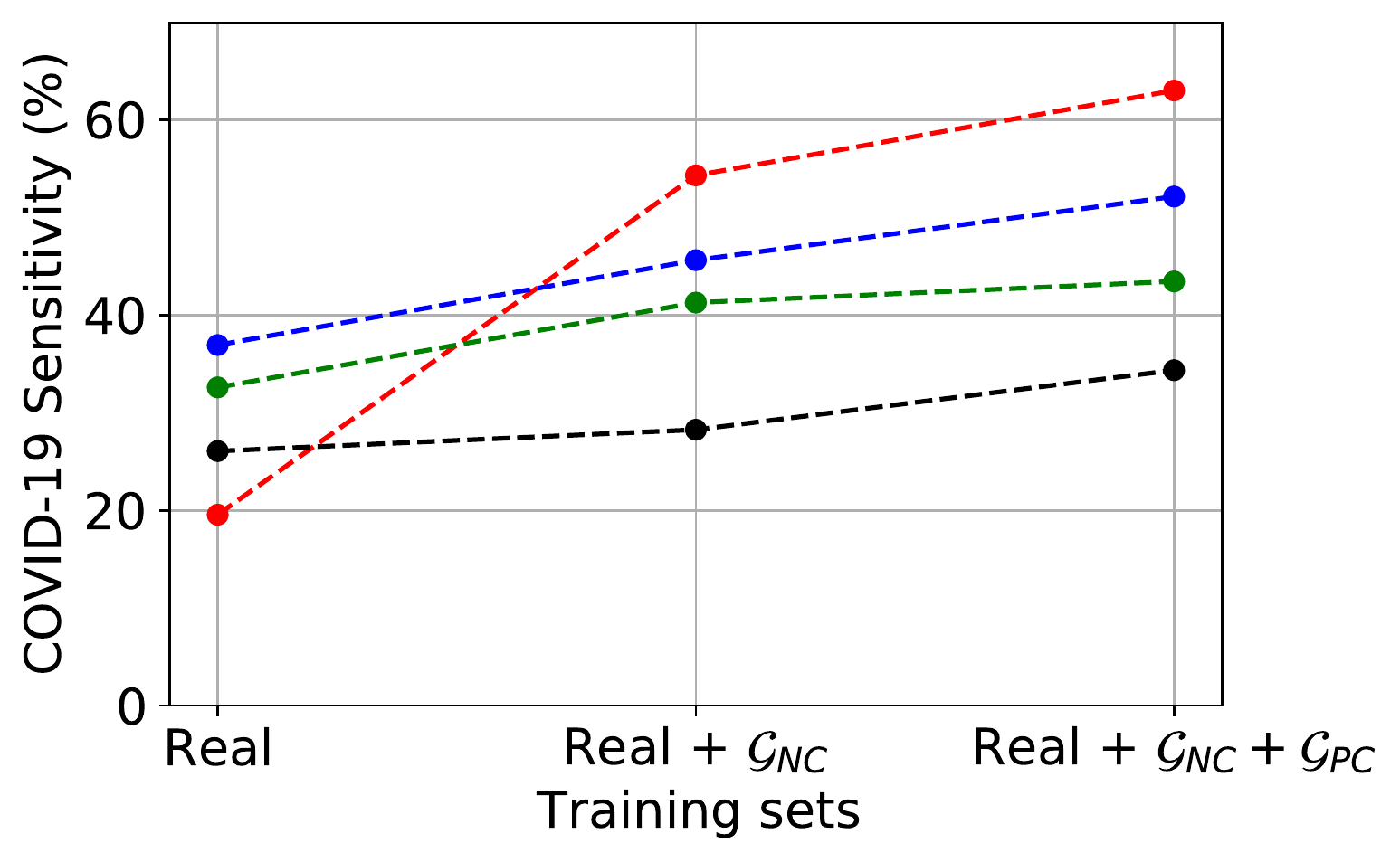} \\
\includegraphics[scale=.45]{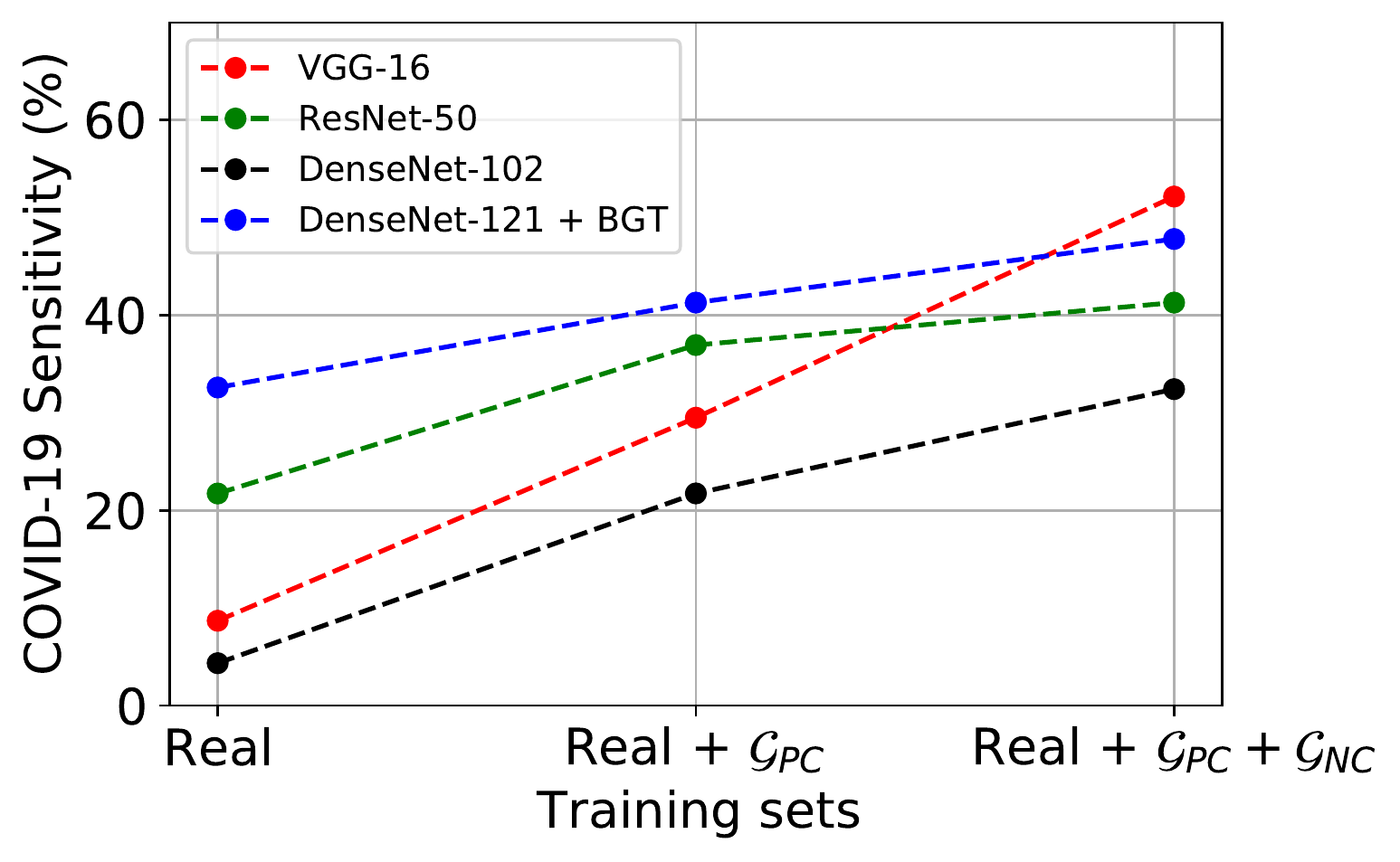}
\end{tabular}
\caption{COVID-19 detection performance results on Normal vs. COVID-19 (top) and Pneumonia vs. COVID-19 (buttom) test sets when trained on real data, and on combined real and synthetic data. For both tasks, synthetic data improves detection performance  of the different classification models.}
\label{fig:tasks}
\end{figure}

We report results in Figure~\ref{fig:tasks} for Normal vs. COVID-19 and Pneumonia vs. COVID-19 tasks when using $\mathcal{G}_{NC}$, instead of $\mathcal{G}_{PC}$, as additional training data across various deep learning architectures. It is evident that for all architectures, adding the synthetic data significantly improves COVID-19 detection performance. For Normal vs. COVID-19, the performance is much better when adding $\mathcal{G}_{NC}$ instead of $\mathcal{G}_{PC}$. We hypothesize that this is due largely to the fact that the number of images in $\mathcal{G}_{NC}$ is much larger.

\begin{figure*}[!htb]
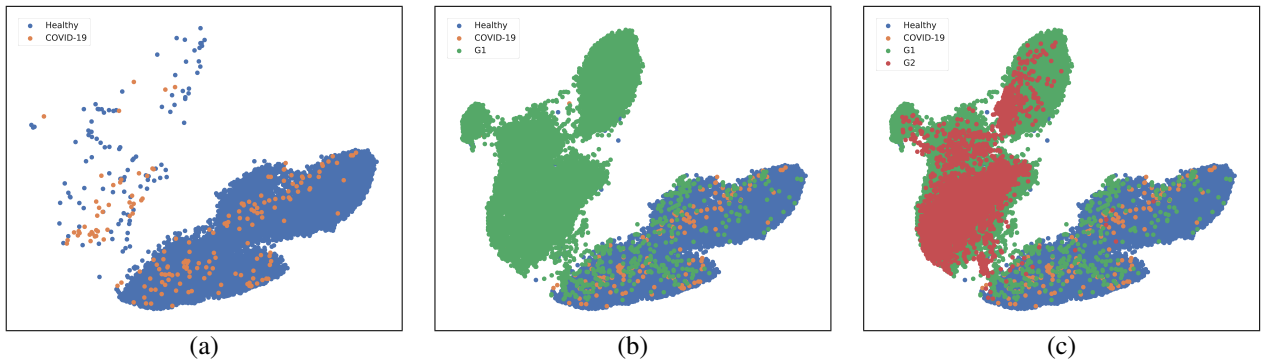

\centering
\begin{tabular}{ccc}
\fbox{\includegraphics[scale=.032]{Figure3a.pdf}} &
\fbox{\includegraphics[scale=.032]{Figure3b.pdf}} &
\fbox{\includegraphics[scale=.032]{Figure3c.pdf}}\\
(a) & (b) & (c)
\end{tabular}
\caption{Two-dimensional UMAP embeddings: (a) Normal vs. COVID-19; (b) Normal vs. COVID-19 + $\mathcal{G}_{NC}$; (c) Normal vs. COVID-19 + $\mathcal{G}_{NC}$ + $\mathcal{G}_{PC}$; Here, G1 and G2 denote $\mathcal{G}_{NC}$ and $\mathcal{G}_{PC}$, respectively.}
\label{fig:umaps}
\end{figure*}

In Figure~\ref{fig:umaps}, we display the two-dimensional Uniform Manifold Approximation and Projection (UMAP) embeddings of the features with the objective of visualizing the difference between the original and synthetic data. Figure~\ref{fig:umaps}(a) shows that the original examples exhibit low interclass variation and consist of outliers. In Figures~\ref{fig:umaps}(b) and 3(c), the synthetic samples are in a different distribution in the feature space. While the UMAP embeddings
may not be interpreted as a justification that the synthetic examples actually consist of COVID-19 symptoms from a clinical perspective, it is, however, important to note that the distribution of the synthetic images is significantly different than that of normal images; thereby enabling a proper decision boundary.

\begin{figure*}[!htb]
\centering
\includegraphics[width=5.5in, height=2.35in]{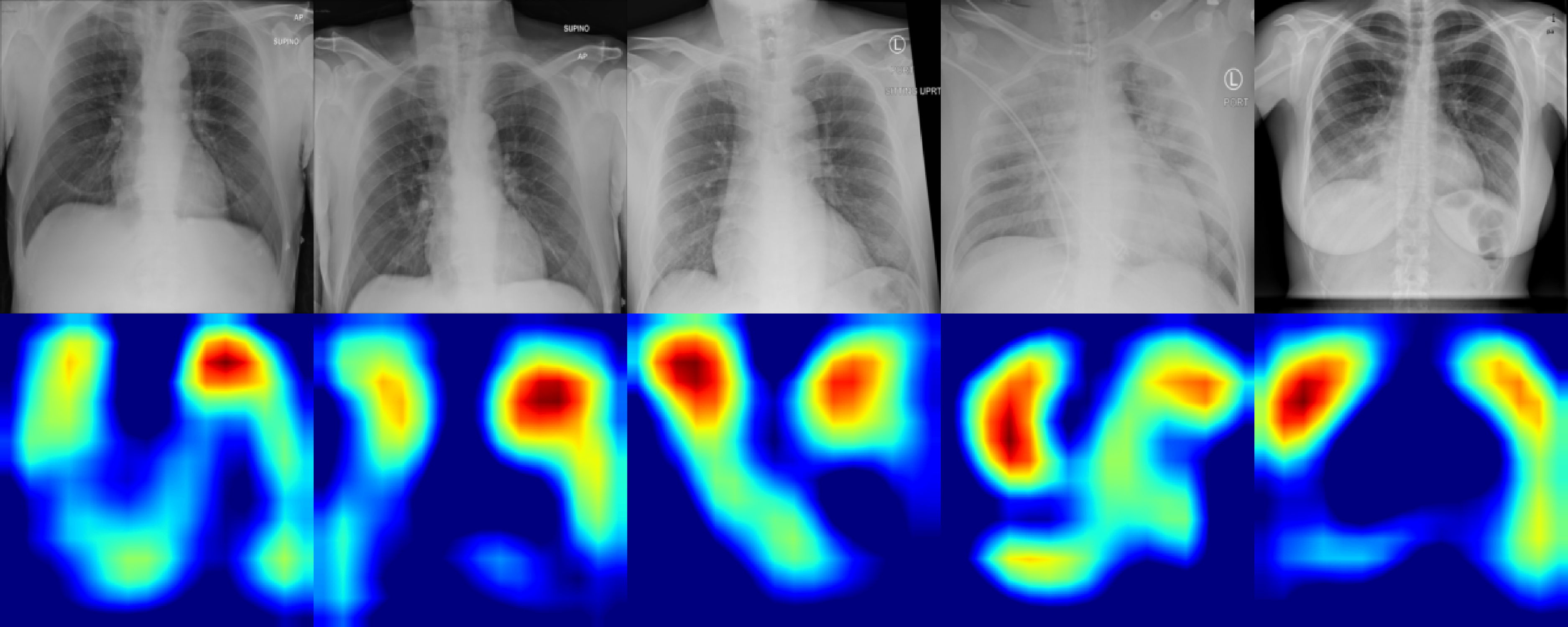}\\[.6ex]
\includegraphics[width=5.5in, height=2.35in]{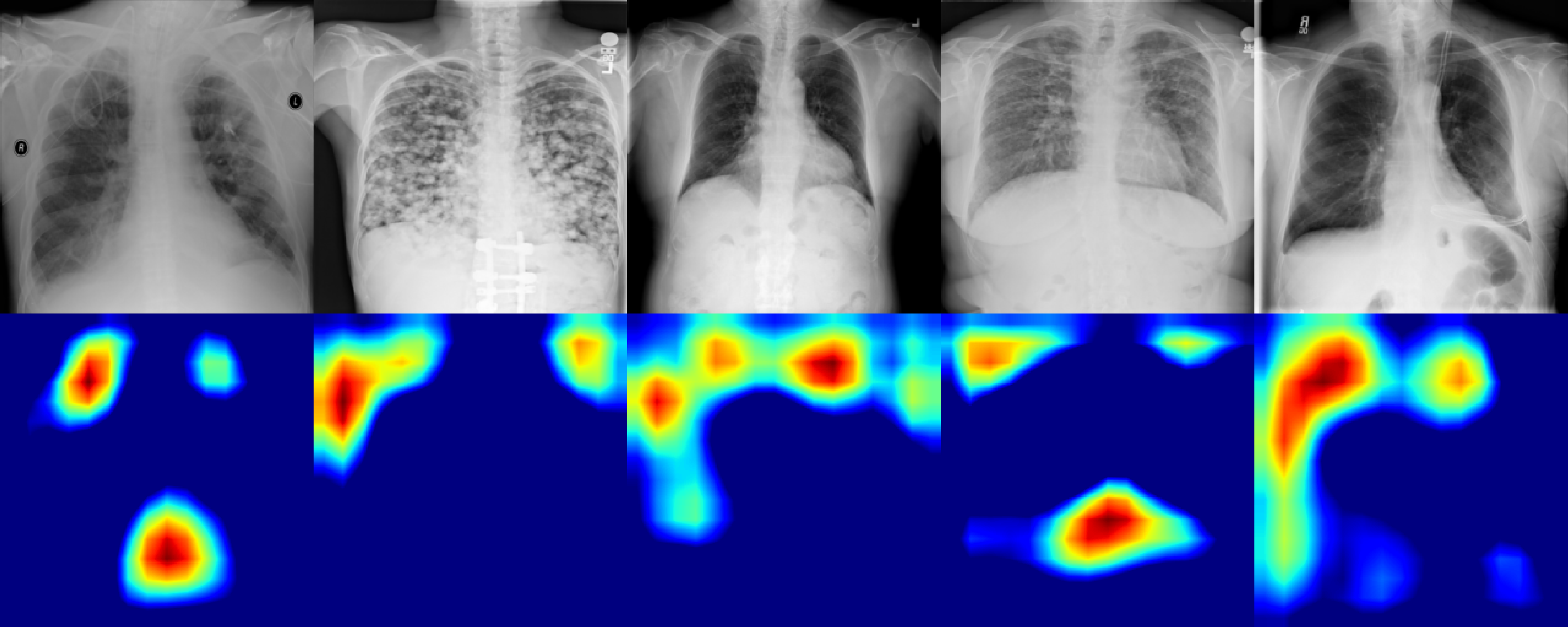}
\caption{Saliency maps for the correctly classified COVID-19 (top two rows) and Non-COVID-19 (bottom two rows) images by the proposed model. Notice that for images that are classified as COVID-19, our model highlights the areas within the lungs, whereas for Non-COVID-19 images, the most important regions are around the lungs.}
\label{Fig:gradcam_xrays}
\end{figure*}

To visually explain the decisions made by the model in the sense that why an X-ray image is classified as COVID/Non-COVID, we use the gradient-weighted class activation map (Grad-CAM) to generate the saliency maps that highlight the most influential features affecting the predictions. Since the convolutional feature maps retain spatial information and that each pixel of the feature map indicates whether the corresponding visual pattern exists in its receptive field, the output from the last convolutional layer of the deep neural network shows the discriminative region in an image. To distinguish between the predicted COVID-19 and Non-COVID-19 images, we visualize the saliency maps for images that are correctly classified as COVID-19 and Non-COVID-19 (normal) by the proposed model. As shown in Figure~\ref{Fig:gradcam_xrays}, the class activation maps for Non-COVID-19 (normal) demonstrate high activations for regions around the lungs, suggesting that there are no prediction features indicating that the disease is present. For most of the images that are correctly classified as COVID-19, the highlighted regions are within the lungs. Notice that in some cases, the model only highlights a specific part of the lung (e.g. left or right), which shows that COVID-19 features are present only on one side.

\section{Conclusion}
\label{Conclusion}
We release a publicly available dataset consisting of 21,295 synthetic COVID-19 CXR images to be used for training machine learning models in computer-aided diagnosis. We find that using these images can alleviate heavy class imbalance problems across multiple deep learning architectures for COVID-19 detection. Salient features also suggest that the distribution of synthetic images are different from other classes, and hence enable a proper decision boundary.




\bibliography{references}
\bibliographystyle{icml2021}





\end{document}